\documentclass[fleqn,usenatbib]{mnras}

\usepackage{newtxtext,newtxmath}

\usepackage[T1]{fontenc}

\DeclareRobustCommand{\VAN}[3]{#2}
\let\VANthebibliography\thebibliography
\def\thebibliography{\DeclareRobustCommand{\VAN}[3]{##3}\VANthebibliography}

\usepackage{graphicx}	            
\usepackage{amsmath}	            
\usepackage{bm}                     
\usepackage{float}
\usepackage{caption}                
\usepackage{subcaption}
\usepackage[dvipsnames]{xcolor}     
\usepackage[normalem]{ulem}         
\usepackage{nicefrac, xfrac}        
\usepackage{makecell}
\usepackage{threeparttable}         

\graphicspath{{Figures/}}

\newcommand{\rlc}{R_\text{LC}}

\title[Pulsar Equation with PINNs]{Solving the Pulsar Equation using Physics-Informed Neural Networks}

\author[P. Stefanou et al.]{
Petros Stefanou,$^{1,2}$\thanks{E-mail: petros.stefanou@uv.es}
Jorge F. Urb\'an,$^{1}$
Jos\'e A. Pons$^{1}$
\\
$^{1}$Departament de Física Aplicada, Universitat d'Alacant, Ap. Correus 99, E-03080 Alacant, Spain\\
$^{2}$Departament d'Astronomia i Astrofísica, Universitat de València, Dr. Moliner 50, E-46100, Burjassot, València, Spain
}

\date{Accepted XXX. Received YYY; in original form ZZZ}

\pubyear{2023}

\begin{document}
\label{firstpage}
\pagerange{\pageref{firstpage}--\pageref{lastpage}}
\maketitle

\begin{abstract}

In this study, Physics-Informed Neural Networks (PINNs) are skilfully applied to explore a diverse range of pulsar magnetospheric models, specifically focusing on axisymmetric cases. The study successfully reproduced various axisymmetric models found in the literature, including those with non-dipolar configurations, while effectively characterizing current sheet features. Energy losses in all studied models were found to exhibit reasonable similarity, differing by no more than a factor of three from the classical dipole case. This research lays the groundwork for a reliable elliptic Partial Differential Equation solver tailored for astrophysical problems. 
Based on these findings, we foresee that the utilization of PINNs will become the most efficient approach in modelling three-dimensional magnetospheres. This methodology shows significant potential and facilitates an effortless generalization, contributing to the advancement of our understanding of pulsar magnetospheres.

\end{abstract}

\begin{keywords}
magnetic fields; pulsars; stars: neutron; 
\end{keywords}



\section{Introduction}

Physics-Informed Neural Networks (PINNs) (\citet{Lagaris_Likas_Fotiadis_97,Raissi_Perdikaris_Karniadakis_19}) is a relatively new but very promising family of PDE solvers based on Machine Learning (ML) techniques. 
This method is suitable to obtain solutions of Partial Differential Equations (PDEs) describing the physical laws of a given system by taking advantage of the very successful modern ML frameworks and incorporating physical knowledge. 
In recent years, PINN-based solvers have been used to solve problems in a great variety of fields: fluid dynamics (\citet{Cai21}), turbulence in supernovae (\citet{Karpov22}), radiative transfer (\citet{Korber23}), black hole spectroscopy (\citet{Luna23}), cosmology (\citet{Chantada23})large scale structure of the universe (\citet{Aragon-Calvo_2019}), galaxy model fitting (\citet{Aragon-Calvo_Carvajal_2020}), inverse problems (\citet{Pakravan_Mistani_Aragon-Calvo_Gibou_2021}) and many more. 
In our previous work (\citet{Urban_Stefanou_Dehman_Pons_23}, Paper I hereafter) we presented a PINN solver for the Grad-Shafranov equation, which describes the magnetosphere of a slowly rotating neutron star endowed with a strong magnetic field (a magnetar) in the axisymmetric case. 
In that paper, we demonstrated the ability of the network to be trained for various boundary conditions and source terms simultaneously.  
In this work, our purpose is to extend our PINN approach to the more general -- and more challenging -- case of rapidly rotating neutron stars (pulsars). 
The rotating case presents new interesting challenges related to the presence of current sheets in the magnetosphere. 
Our implementation is able to deal with these pathological regions sufficiently well, demonstrating its potential for problems where classical methods struggle. 

Pulsar magnetospheres have been studied extensively in the last 25 years with various approaches, each with its advantages and limitations. 
\citet{Contopoulos_Kazanas_Fendt_99} were the first to solve the axisymmetric, time~-independent problem, a result that was later confirmed and improved by \citet{Gruzinov_2005} and \citet{Timokhin_06}. \citet{Spitkovsky06} used a full MHD time-dependent code and was able to acquire solutions for aligned and oblique rotators. 
Solutions for arbitrary inclination were also obtained by \citet{Petri12} who used a time~-dependent pseudo-spectral code. 
More recent approaches involve large-scale Particle-in-Cell simulations that include the influence of accelerated particles (\citet{Cerutti15, Philippov18}).
All of these, and many related, works have improved our understanding of the pulsar magnetosphere.
However, some important questions remain still without an answer.

The structure of the paper is the following. After a brief summary of the relevant equations and boundary conditions in section \S2, we will describe the PINN method in \S3, with emphasis in the new relevant details with respect to the magnetar problem (paper I). In section \S4 we present our results, showing that
we are able to reproduce the well-established axisymmetric results encountered in the literature, but also indicating that new, possibly unexplored solutions can be encountered. We summarize our most relevant conclusions and discuss possible improvements and future extensions in section \S5.

\section{Pulsar magnetospheres.}

Our aim is to extend our previous results from paper I to find numerical solutions of the pulsar equation, that can be written as follows:
\begin{equation}\label{eq:relativistic_grad_shafranov}
    \bm{\nabla} \times \left(\bm{B} - \beta^2 \bm{B}_p\right) = \alpha \bm{B}, 
\end{equation}
where $\beta = \nicefrac{v}{c} = \nicefrac{|\bm{\Omega} \times \bm{r}|}{c} = \nicefrac{\varpi}{\rlc}$ is the co-rotational speed in units of the speed of light, with $\rlc = \nicefrac{c}{\Omega}$ being the light-cylinder (LC) radius and $\bm{B}_p = \bm{B} - \bm{B}_\phi$ is the magnetic field perpendicular to the direction of rotation. Here, $\alpha$ is a scalar function given by
\begin{equation} \label{eq:alpha}
    \alpha = \frac{4 \pi}{c}
    \left(\bm{J}-\rho_e \bm{v}\right)\cdot \frac{\bm{B}}{B^2}.
\end{equation}
representing the ratio between the field-aligned component of the current in the corotating frame and the local magnetic field strength. We refer to the comprehensive and thorough recent review by \citet{PK2022} (and references therein) for a complete historical overview of magnetospheric physics and the mathematical derivation of the equations.

As in Paper I, we focus in the axisymmetric case and, for convenience, we use compactified spherical coordinates $(q, \mu, \phi)$ where $q=\nicefrac{1}{r}, \mu = \cos{\theta}$. 
In these coordinates, any axisymmetric magnetic field can be written in terms of a poloidal and a toroidal scalar stream function $P$ and $T$ as 
\begin{equation}\label{eq:B_axisymmetric}
    \bm{B} = \frac{q}{\sqrt{1-\mu^2}} \left(\bm{\nabla} P \times \hat{\phi} + T \hat{\phi}\right). 
\end{equation}
Here $P$ is related to the magnetic flux and is constant along magnetic field lines.
Plugging this expression into Eq.~\eqref{eq:relativistic_grad_shafranov} and taking the toroidal component, we get
\begin{equation}\label{eq:gradP_gradT}
    \bm{\nabla} P \times \bm{\nabla} T = 0,
\end{equation}
which means that $T$ is only a function of $P$ ($T=T(P)$) and, therefore, is also constant along magnetic field lines. 
The poloidal component of Eq.~\eqref{eq:relativistic_grad_shafranov} gives the well-known Pulsar Equation (\citet{Michel_73, Scharlemann_Wagoner_73}, which in our coordinates reads
\begin{equation}\label{eq:pulsar_equation}
    \left( 1 - \beta^2 \right) \Delta_{\text{GS}} P + 2 \beta^2 q^2 \left( q \partial_q P + \mu \partial_\mu P \right) + G(P) = 0,
\end{equation}
where $G(P) = T T'$ and $\Delta_{\text{GS}}$ is the Grad-Shafranov operator, given by
\begin{equation} \label{eq:grad_shafranov_operator}
    \Delta_{\text{GS}} \equiv q^2 \partial_q \left(q^2 \partial_q \right) + \left(1-\mu^2 \right)q^2 \partial_{\mu \mu}. 
\end{equation}
Notice that in the limiting case where $\beta = 0$, we ignore the rotationally induced electric field and Eq. \eqref{eq:pulsar_equation} is reduced to the Grad-Shafranov equation.
Hereafter we will use the following shorthand notation for the extended operator
\begin{equation}
        \Delta_{GS\beta} \equiv \left( 1 - \beta^2 \right) \Delta_{\text{GS}} + 2 \beta^2 q^2 \left( q \partial_q + \mu \partial_\mu \right) ~.
\end{equation}

A crucial difference in the modelling of magnetar and pulsar magnetospheres is the source of poloidal currents $T T'$. 
In magnetars, currents are assumed to be injected into the magnetosphere due to the magnetic field evolution in the NS's crust, which twists the magnetosphere \citep{Akgun2018}.
Therefore, the source term in Eq.~\eqref{eq:pulsar_equation} is given either as a user-parametrised model (as for example in \citet{Akgun_Miralles_Pons_Cerda_16} for the 2D problem or in \citet{Stefanou_Pons_Cerda_23} for the 3D problem) or in a self-consistent manner by coupling the interior evolution with the magnetosphere. 
We adopted the latter approach in Paper I (see the astrophysical application in section 5 of that paper), where a series of magnetospheric steady-state solutions were obtained for each time-step of the internal magneto-thermal evolution. 

In the pulsar magnetosphere, however, the source of current is the LC. 
Lines that cross the LC have to open up and bend, giving rise to azimuthial fields and currents.
We refer to the region with open field lines as the open region.
On the other hand, lines that do not cross the LC, but turn back to the surface, co-rotate rigidly with the the star. We refer to this region as the closed region.
The source term $TT'$ must be determined self-consistently to ensure smooth crossing of the lines along the LC. 
In particular, at the LC where $\beta = 1$, Eq.~\eqref{eq:pulsar_equation} takes the simple form 
\begin{equation}
    -q^2\left(q \partial_q P + \mu \partial_\mu P\right) = 2 B_z = TT',
\end{equation}
which places a constraint on the possible functions $T (P)$. 
This makes Eq.~\eqref{eq:pulsar_equation} complicated to solve, as two functions have to be determined simultaneously.
In order to do so, additional physical boundary conditions have to be imposed (see next subsection).

Through this work, we measure distances in units of the radius of the star $R$ and magnetic fields in units of the surface magnetic field strength at the equator $B_0$ (note that the surface magnetic field strength at the poles is $2 B_0$).
In these units, the magnetic flux is measured in units of the total poloidal flux $P_{0} = B_0 R^2$.
It is convenient to additionally define the total magnetic flux carried by field lines that cross the LC in a non-rotating dipole $P_1 = P_0 (\nicefrac{R}{\rlc})$, which will be useful in what follows.
Finally, the toroidal stream function is measured in units of $B_0 R$.

\subsection{Boundary Conditions}

We assume that the magnetic field at the surface of the star $P(q=1)$ is knnown. 
For example, in the case of a dipole
\begin{equation}\label{eq:bc_sufrace}
    P (q=1, \mu) = B_0 (1 - \mu^2)
\end{equation}
but we will also explore other options. 
The last closed field line, called separatrix, will be labelled by $P = P_c$. 
It marks the border between open and closed (current-free) regions. The value $P_c$
corresponds to the total magnetic flux that crosses the light cylinder.
The point where this line meets the equator is the called Y-point and it should lie at a radius $r_c$, somewhere between the surface and the LC (see e.g. \citet{Timokhin_06} for a detailed discussion).
Far away from the surface, field lines should become radial, resembling a split monopole configuration \citep{Michel_73}. Inside the closed region, the magnetosphere is current-free and purely poloidal. No toroidal fields are developed, so that
\begin{equation}\label{eq:bc_separatrix}
    T (P > P_c) = 0
\end{equation}
by definition.

In the classical model, a current sheet should develop along the separatrix, supported by the discontinuity of the toroidal magnetic field between closed ($B_\phi = 0$) and open ($B_\phi \neq 0$) regions.
Another current sheet should exist along the equator and beyond the Y-point, supported by the opposed directions of the magnetic field lines between the two hemispheres.
The current sheet forms the return current that accounts for the current supported by out-flowing particles along the open field lines and closes the current circuit. 

All this motivated the seminal works to impose equatorial symmetry by defining a numerical domain only on one hemisphere and imposing another boundary condition $P=P_c$ at the equator. To reproduce these models, we will impose that 
along the equator and beyond the Y-point
\begin{equation}\label{eq:bc_equator}
    P (q < q_c, \mu = 0) = P_c.
\end{equation}
Here $q_c = \nicefrac{1}{r_c}$ is the corresponding location of the Y-point in compactified coordinates.
However, later works have relaxed this assumption \citep{Contopoulos_2014}, proposing a more general solution without a separatrix and with a transition region close to the equator. We will explore both cases in our results section.

\subsection{Total Energy and Radiated Power}

The total energy of the electromagnetic field in the magnetosphere is given by
\begin{equation}
    {\cal E} = \frac{1}{8 \pi}\int (B^2 + E^2) dV.
\end{equation}
It is convenient to describe the electromagnetic energy content of a given model in terms of the excess energy of a particular magnetospheric solution with respect to the non-rotating dipole, 
\begin{equation}\label{eq:excess_energy}
    {\Delta \cal E} = \frac{{\cal E} - {\cal E}_d}{{\cal E}_d},
\end{equation}
where
\begin{equation}\label{eq:dipole_energy}
    {\cal E}_d = \frac{1}{3} B_0^2 R^3
\end{equation}
is the total magnetic energy of a non-rotating dipole.

The total power radiated away by a rotating magnetosphere can be calculated by integrating the Poynting flux over a sphere far away from the star
\begin{equation}\label{eq:luminocity}
    {\cal \dot{E}} = \frac{c}{4 \pi} \int (\bm{E} \times \bm{B}) \cdot \hat{\bm{r}} d \omega = -\frac{c}{\rlc} \int_0^{P_c} T dP,
\end{equation}
where $\omega$ is the solid angle. Again, it is convenient to measure the relative difference of the radiated power with respect to the classical 
order of magnitude estimate
\begin{equation}
    {\cal \dot{E}}_d = \frac{B_0^2 R^6}{\rlc^4}c,    
\end{equation}
and hereafter we will express ${\cal \dot{E}}$ in units of ${\cal \dot{E}}_d$.
\footnote{We should stress that the (unfortunately) often used $\sin^2\theta$ dependence with the inclination angle in ${\cal \dot{E}}_d$ is unphysical. It misleads to conclude that an aligned rotator does not emit, but it actually does radiate in a similar amount (but a factor of about 2) than the orthogonal rotator.}
\begin{figure*}
     \centering
    \includegraphics[width=0.9\textwidth]{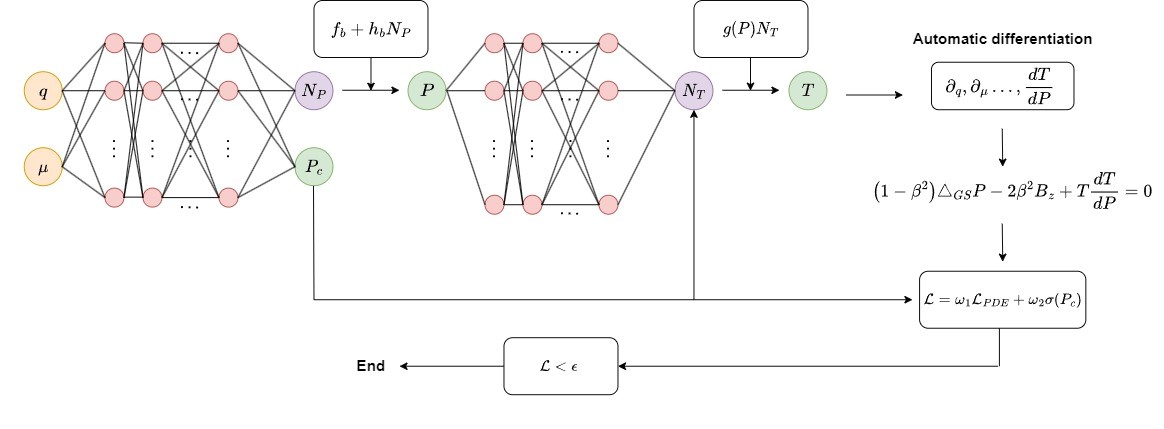}
    \caption{A sketch of the network structure. Two sub-networks are employed to ensure that $P = P (q, \mu)$ and $T = T (P)$.}
    \label{fig:network_structure}
\end{figure*}
\section{Network structure and training algorithm}

In Paper I, we used a PINN to calculate approximate solutions of the axisymmetric Grad-Shafranov equation. 
We refer the reader to that paper for a more detailed description of the generic implementation.
In this section, we will briefly summarise the main points and highlight the novelties introduced to adapt our solver to the pulsar case. The principal changes introduced aim at enforcing the physical constraint (boundary conditions, or the $T(P)$ requirement) by construction instead of leaving the job to minimize additional terms in the loss function, which usually does not reach the required accuracy.

We consider solutions in points $(q, \mu) \in \mathcal{D}$ in a 2-dimensional domain $\mathcal{D}$. 
We denote by $\partial \mathcal{D}$ the boundary of this domain. \
To account for the equatorial constraint in Eq. \eqref{eq:bc_equator}, we  will consider the equatorial line beyond $r_c$ as part of $\partial \mathcal{D}$.
In order to ensure that $P$ depends on the coordinates while $T$ is solely a function of $P$, we design a network structure consisting of two sub-networks that are trained simultaneously. 
The output of each sub-network depends only on its corresponding input

The first sub-network takes the coordinates $(q, \mu)$ as input and returns as output a function that we denote by $N_P$. 
Then, $P$ is calculated using 
\begin{equation} \label{eq:P_parametrisation}
    P (q, \mu) = f_b (q, \mu) + h_b (q, \mu) N_P (q, \mu; \Theta).
\end{equation}
where  $f_b$ can be any function in $\mathcal{D}$ that satisfies the corresponding
boundary conditions at $\partial \mathcal{D}$.
$h_b$ is an arbitrary function representing some measure of the distance to the boundary, that must vanish at $\partial \mathcal{D}$.
Both user-supplied functions $f_b$ and $h_b$ depend only on the coordinates and are unaffected by the PINN.
There is some freedom to decide their specific form, as long as they retain certain properties (e.g. they are sufficiently smooth) and they have the desired behaviour at the boundary.
The only part that is adapted during training is $N_P$ through its dependence on the trainable parameters $\Theta$.
With this \textit{parametrisation} (or \textit{hard enforcement}) approach we ensure that boundary conditions are exactly satisfied by construction. This differs from the other usual approach, consisting of adding more terms related to the boundary conditions in the loss function. We will specify the particular form of the functions $f_b (q, \mu)$ and 
$h_b (q, \mu)$ in the next section when we discuss different cases.

Next, the second sub-network takes $P$ as input and returns $N_T$ as output.
This automatically enforces that $T=T(P)$, required by  Eq.~\eqref{eq:gradP_gradT}. The new network output 
$N_T (P; \Theta)$ is used to construct the function $T(P)$ as follows:
 \begin{equation} \label{eq:T_parametrisation}
    T (P) = g (P) N_T (P; \Theta)~,
\end{equation}
where $g$ is another user-supplied function to include other physical restrictions. For example, we can use a step function to treat a possible discontinuity of $T$ at the border between open and closed regions (Eq.~\eqref{eq:bc_separatrix}). 
Another advantage of this approach is that $T'$ is calculated from $T$ via automatic differentiation in the second sub-network.

A subtle point in the pulsar problem is the determination of the critical value $P_c$ that separates the 
open and closed regions.  We have explored two different approaches: \textbf{(a)} $P_c$ is self-determined by the network and \textbf{(b)} $P_c$ is fixed to some value.
In the first case, $P_c$ is considered an extra output of the first sub-network, closely connected to $P$, resembling the approach taken by classical solvers for this problem.
In contrast, the second method involves fixing $P_c$ as a hyperparameter to a predetermined value, leaving the network to discover a solution that aligns with this fixed value. This introduces the necessity to loosen some of the other imposed constraints, like the position of the Y-point, and encourages the exploration of solution properties from a fresh standpoint.
\bigskip 

During the network training process, both $P$ and $T$ (and possibly $P_c$, if we follow the case \textbf{(a)} above) are obtained simultaneously by minimising the following loss function
\begin{equation}
    \mathcal{L} = \omega_1 \mathcal{L}_{PDE} + \omega_2 \sigma(P_c),\label{eq:loss}
\end{equation}
where 
\begin{equation}
     {\cal L}_{PDE} = \frac{1}{N} \sum_{(q, \mu) \in {\cal D}} {\left[ \Delta_{GS\beta} P (q, \mu)+ G (P (q, \mu)) \right]^2} ~.
\end{equation}
Here, $N$ represents the size of the training set, and $\omega_1$ and $\omega_2$ are adjustable parameters. In approach \textbf{(b)}, $\omega_2 = 0$, whereas in approach \textbf{(a)}, $\omega_2$ is adjustable. 
Additionally, $\sigma$ denotes the standard deviation of the $P_c$ values for the points in the training set. 
$P_c$ is a global magnetospheric value that should not depend on the coordinates. However, during training, there is no guarantee that this value will be the same for all the points that are considered. To ensure that the values of $P_c$ at arbitrary points are as close to each other as possible, we minimise the standard deviation ($\sigma(P_c)$) of the values of $P_c$ at all the points of the training set by including an additional term in the loss function \eqref{eq:loss}.
This inclusion guarantees that the network's output for $P_c$ remains constant and independent of the coordinates $(q, \mu)$. 
A schematic representation of our network's structure can be found in Fig. \ref{fig:network_structure}.

We consider a fully-connected architecture for both sub-networks consisting of 4 hidden layers for the first one and 2 hidden layers for the second one. All layers have 40 neurons.
Our training set consists of $N=5000$ random points $(q, \mu) \in \mathcal{D}$, which are periodically changed every 500 epochs in order to feed the network with as many distinct points as possible.
The total number of epochs is 35000. This number may seem big at first sight, but is necessary because of the amount of points considered.
In order to minimise the loss function, we use the ADAM optimisation algorithm \citep{2014arXiv1412.6980K} with an exponential learning rate decay. 
We have also considered in this work the idea of introducing trainable activation functions, suggested in PINNs originally by \citealt{2020JCoPh.40409136J}. 
For each hidden layer $k$, the linear transformation performed at that layer is multiplied by a trainable parameter $c_k$. We found that this practice can accelerate convergence but a more rigorous study is out of the scope of this paper.

\begin{figure}
    \centering
    \includegraphics[width=\columnwidth]{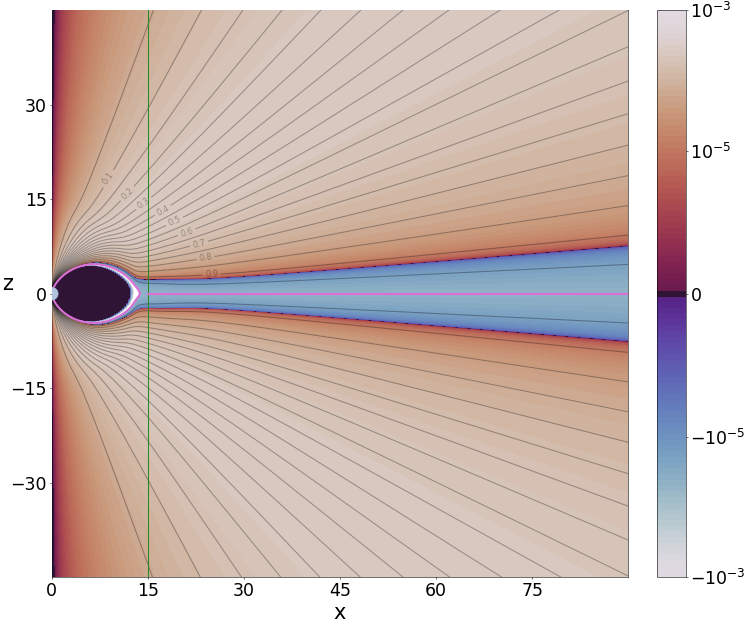}
    \caption{The classical axisymmetric pulsar magnetosphere. Fade black lines: magnetic field lines as contours of $P$. Thick pink line: separatrix and equatorial current sheet, where $P = P_c$. Vertical green line: light cylinder. Colourmap: the source current $G(P)$. Labels on the field lines are in units of $P_c$. Colourbar is in symmetrical logarithmic scale. z and x axes are in units of the stellar radius $R$. The bulk of the return current flows along the current sheet, with a small percentage flowing along open field lines}
    \label{fig:G_colourmap_classical}
\end{figure}

\section{Magnetospheric models}

In this section, we present the results obtained for various cases and under different physical assumptions. Our analysis encompasses the successful reproduction of all distinct axisymmetric models found in 
literature, which include:
\begin{enumerate}
    \item The classical solutions \citep{Contopoulos_Kazanas_Fendt_99, Gruzinov_2005},
    \item The family of solutions with varying locations of the Y point \citep{Timokhin_06},
    \item The improved solution where the separatrix current sheet is smoothed out \citep{Contopoulos_2014},
    \item The non-dipolar solutions \citep{Gralla_2016}.
\end{enumerate}
Each of these models presents unique challenges and characteristics, and we will delve into the outcomes achieved for each one. In particular, the overall magnetospheric configuration, the poloidal flux at the separatrix $P_c$, the functions $T (P)$ and $G(P)$, and the energy losses $\dot{\cal E}$ agree with the previous works. 

\begin{figure}
    \centering
    \includegraphics[width=\columnwidth]{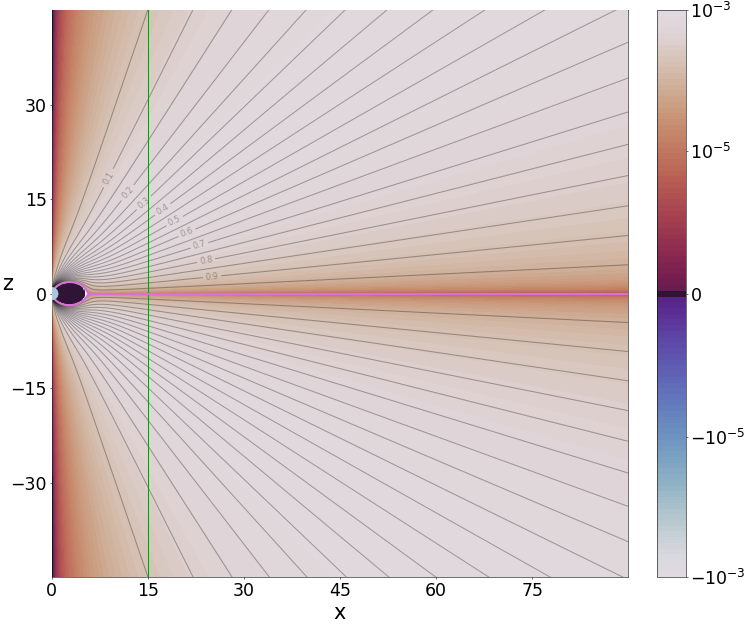}
    \caption{Same as figure \ref{fig:G_colourmap_classical} but with the Y-point positioned at $r_c = 0.4 \rlc$. The region where the return current flows trhrough open field lines in negligible.}
    \label{fig:G_colourmap_classical_ypoint}
\end{figure}

\begin{figure*}
    \begin{subfigure}[t]{\columnwidth}
        \centering
        \caption{}
        \includegraphics[width=\columnwidth]{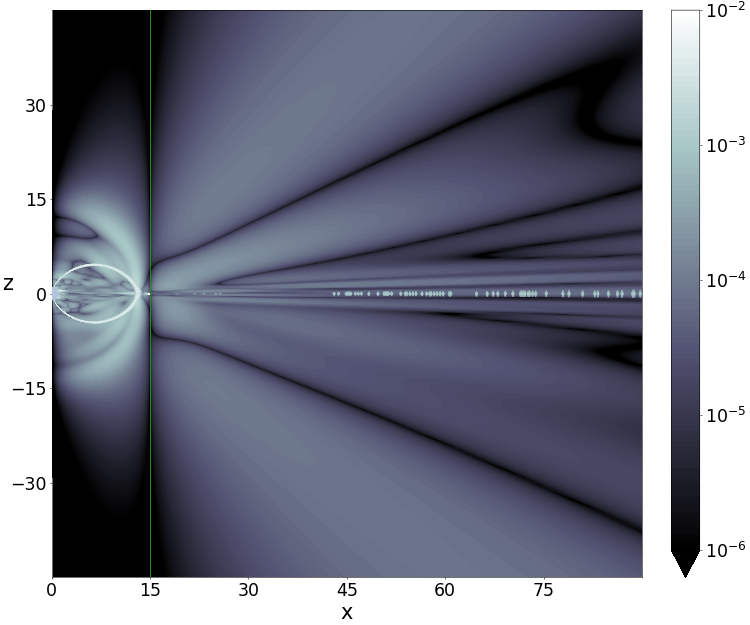}
        \label{fig:pulsar_eq_colourmap} 
    \end{subfigure}
    \begin{subfigure}[t]{\columnwidth}
        \centering
        \caption{}
        \includegraphics[width=\columnwidth]{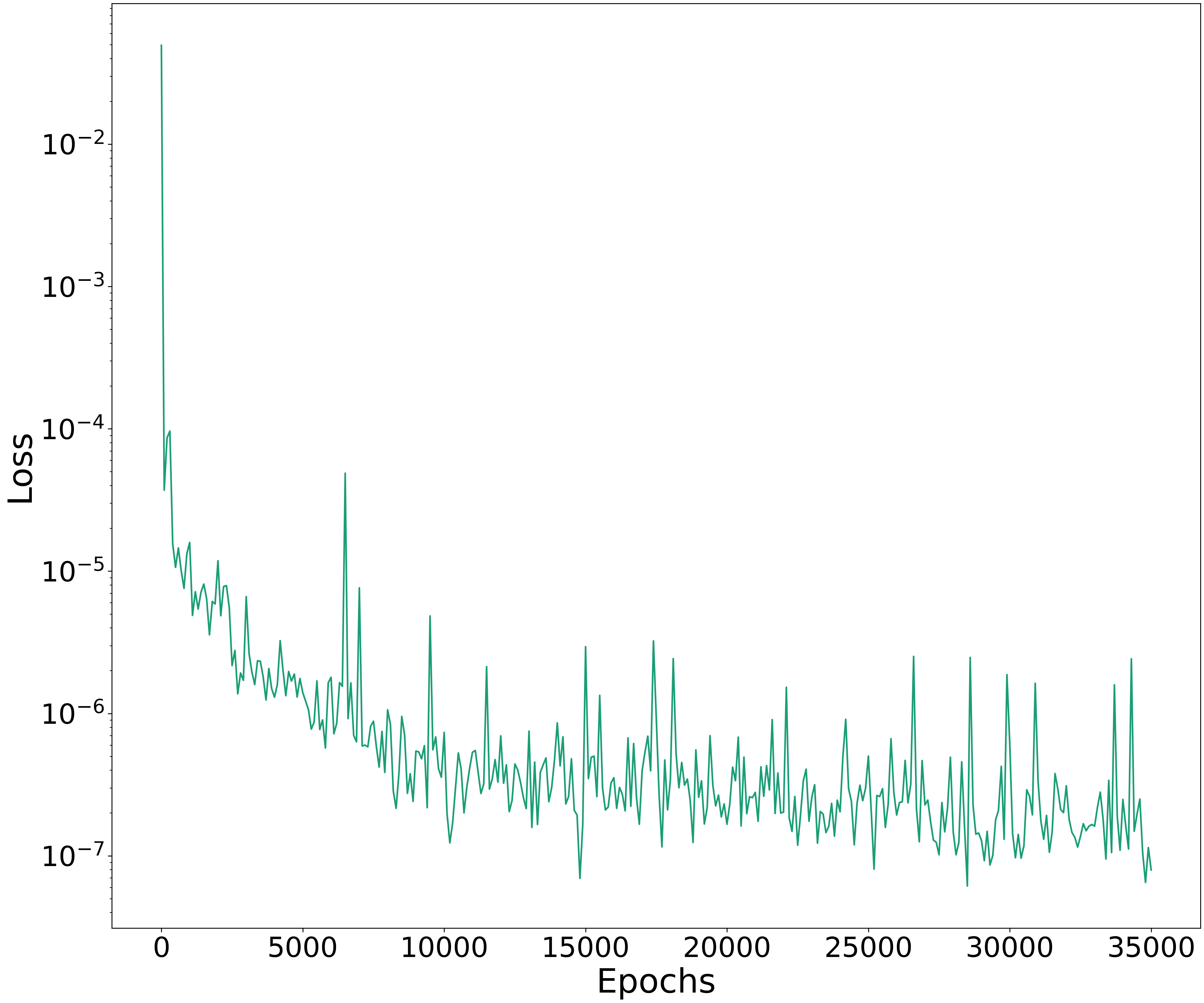}
        \label{fig:loss_vs_epochs}
    \end{subfigure}
    \caption{ \textbf{(a)} Colourmap of the residuals of the pulsar equation. Colourbar is in log scale. \textbf{(b)} Evolution of the loss function with the training epochs. Only the value of the loss every 100 epochs is plotted for clarity. Big spikes correspond to changes in the training set. Small fluctuations can be interpreted as the variance of the PDE residual.}
    \label{fig:reference_solution} 
\end{figure*}

\subsection{Hard enforcement of boundary conditions.}
As discussed earlier, we construct $P$ according to Eq.~\eqref{eq:P_parametrisation} to fulfill the boundary conditions. 
For the cases (i) and (ii) we have employed an $f_b$ with the following form:
\begin{equation}\label{eq:fb_parametrisation_classical}
    f_b (q, \mu) = (1 - \mu^2) \left[ P_c + q(1 - P_c) \frac{\text{ReLU}^3 \left(1 - \frac{q_c}{q}\right)}{ (1 - q_c)^3} \right],
\end{equation}
where ReLU (x) is the \textit{Rectified Linear Unit} function, which returns zero if $x<0$ and $x$ if $x>0$\footnotemark, $q_c = \nicefrac{1}{r_c}$ is the position of the Y-point and $n$ is a free positive parameter.

\footnotetext{Alternative definitions of $\text{ReLU} (x)$ are $x \mathcal{H} (x)$ ($\mathcal{H}$ being the Heaviside step function) and $\text{max} (0,x)$.}

Furthermore, we have chosen the following $h_b$ function:
\begin{equation}
    h_b (q, \mu) = (1 - \mu^2) (1 - q)\sqrt{\text{ReLU}^3 \left(1 - \frac{q_c}{q}\right) + \mu^2},
\end{equation}
where the reader can check that $h_b$ is zero at the boundary. 
The terms with the $\text{ReLU}$ functions enforce that $P = P_c$ at the equator when $q < q_c$, and the third power ensures that $P$ and its first and second derivatives are all continuous.

The parametrisation for $T(P)$ requires the additional function $g$. We use a gaussian\footnotemark transition beginning at $P = P_c$:
\begin{equation}
    g (P) =
    \begin{cases}
        1 & P \leq P_c \\
        e^{- \frac{(P - P_c )^2}{2 (\delta P)^2}} & P>P_c
    \end{cases}
    \label{eq:current_sheet}
\end{equation}
where $\delta P$ is a small number that controls the width of the current sheet, where the transition of T from a finite value to zero takes place. 
Note that $g$ and its first derivative are both continuous at the separatrix, so $T$ and $T'$ are well defined.

\footnotetext{Any other function that asymptotically approaches the Heaviside step function could be used instead.}

For the cases (iii) and (iv), the condition $P = P_c$ at the equator is lifted and consequently, the parametrisation functions must be modified. 
Our choice for these cases is:
\begin{align}
    &f_b (q, \mu) = f_1 (q)  \sum_{l=1}^{l_{\text{max}}} \frac{b_{l}}{l} P_{l}'(\mu) + f_2 (q)  (1 - |\mu|) \label{eq:fb_parametrisation_free_equator} \\
    &h_b (q, \mu) = q  (1-q)  (1 - \mu^2) \label{eq:hb_parametrisation_free_equator} \\
    &g (P) = - P  ~\left( \text{ReLU} \left(1 - \frac{|P|}{P_c}\right) \right) \label{eq:g_parametrisation_free_equator}.
\end{align}

Adjustable weighting functions, denoted as $f_1$ and $f_2$, are utilized to control the relative importance of the two terms. To be specific, $f_1$ should be significant near the surface but diminish at considerable distances from the star, while $f_2$ should vanish close to the surface but be dominant at $q\ll 1$. At intermediate distances, approximately around the LC, both $f_1$ and $f_2$ should be much smaller than $h_b$ to ensure that the neural network contribution in Eq. \eqref{eq:P_parametrisation} dominates.

In order to allow for more versatile configurations beyond the standard dipole representation, the surface boundary condition is presented as a linear combination of magnetic multipoles. To impose a current-free region, the function $g$ is employed, and while it does not necessarily indicate a current sheet, it is chosen to be at least quadratic. This ensures that $T'$ and consequently $G$ experience at least one change of sign, allowing the current circuit to close.

\subsection{Classical solutions}

The solutions obtained in \citet{Contopoulos_Kazanas_Fendt_99, Gruzinov_2005} will be referred to as the \textit{classical solutions}. Fig. \ref{fig:G_colourmap_classical} illustrates contour lines of the poloidal flux function $P$, with the colormap representing the source current $G = T T'$. All the well-known characteristics of the pulsar magnetosphere are observed in these solutions: The open field lines extend beyond the light cylinder and stretch towards infinity, eventually adopting a split monopole configuration at considerable distances, and a current sheet forms at the equator due to the magnetic field's reversal between the north and south hemispheres.

In the region where field lines have $P > P_c$, both the toroidal magnetic field and poloidal current are zero, except for the narrow transition zone $[P_c, P_c + \delta P]$ located just inside the separatrix.
Beyond the separatrix, the blue region illustrates the small portion of the return current that flows back to the star along open field lines. In contrast, the nearly white region just inside the separatrix represents the substantial portion of the return current that flows along the current sheet.
We summarise in Tab. \ref{tab:reference_solution_parameters} the model parameters of different solutions. The first line corresponds to the classical model depicted in Fig. \ref{fig:G_colourmap_classical}.


To showcase the accuracy of our findings, we present in Fig. \ref{fig:pulsar_eq_colourmap} how well our solution aligns with the pulsar equation \eqref{eq:pulsar_equation}. The color map illustrates the absolute error of the pulsar equation for our model, revealing remarkably low values within the bulk of the domain ($\lesssim 10^{-5}$). 
As expected, the error is slightly larger in proximity to the separatrix, a region of discontinuity. Nonetheless, this discrepancy does not impede our solver from effectively approximating the solution throughout the rest of the domain, nor does it significantly affect the solution far from these regions.

In general, we anticipate the maximum error to be of the order of $\sim \sqrt{\cal L}$ since it corresponds to the error of the PDE for a considerable set of random points. Indeed, as depicted in Fig. \ref{fig:loss_vs_epochs}, at the conclusion of the training process, the loss reaches values around $\sim 10^{-7},$ confirming this assumption.
The prominent spikes observed in the figure correspond to the periodic changes of the training set of points. However, it is noticeable that as the training epochs progress, the spikes diminish, indicating that the network has learned to generalize to new points without compromising accuracy.
The small fluctuations between the spikes, should be interpreted as the variance in the approximation to the solution.
As the network adapts its parameters to acquire the solution, it cannot simultaneously reconcile all the training points. Consequently, it fluctuates around a mean instead of finding a single minimum value.

With the same PINN, 
it is straightforward to produce solutions with different positions of the Y-point, as in \citet{Timokhin_06} simply by varying the parameter $q_c$ in Eq.~\eqref{eq:fb_parametrisation_classical}.
Fig. \ref{fig:G_colourmap_classical_ypoint} shows an example of such a solution, where the Y-point lies at a distance $r_c = 0.4 \rlc$. In this case, the totality of the return current flows through the current sheet. Interestingly, we observe that 
the luminosity for $r_c = 0.4 \rlc$ is an order of magnitude larger than the classical solution (see Table \ref{tab:reference_solution_parameters}).

As discussed in \citet{Timokhin_06}, both the luminosity and the  total energy
stored in the magnetosphere increase with decreasing $r_c$, so the magnetosphere will generally try to achieve the configuration with the minimum energy, that is, with $r_c$ as close as possible to the light cylinder. Therefore, although the configurations with a small 
$r_c$ are mathematically sound and very interesting from the astrophysical point of view (much larger luminosity), they are probably short-lived and less frequent in nature than the standard configuration.

\begin{figure}
    \centering
    \includegraphics[width=\columnwidth]{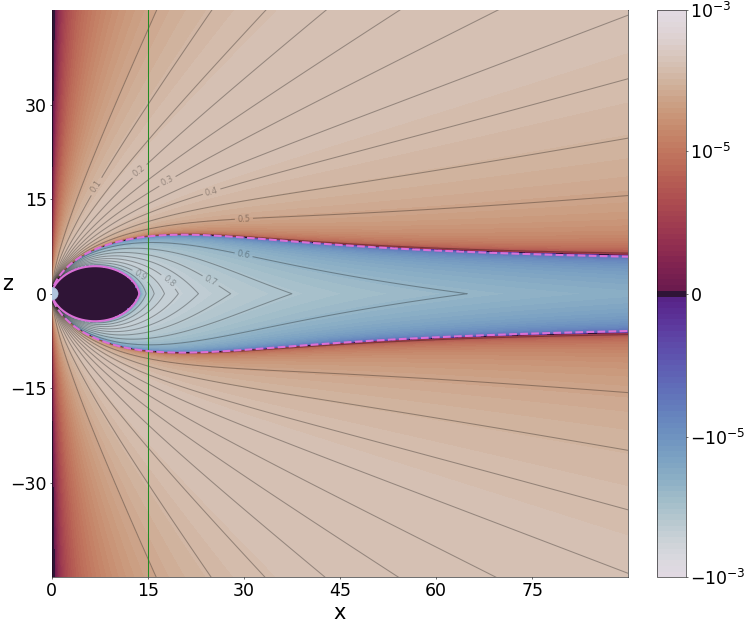}
    \caption{Same as figure \ref{fig:G_colourmap_classical} but for the case of non-fixed equatorial boundary condition. Field lines that cross the light cylinder can close through the current sheet. The solid pink line indicates the line $P = P_c = 1.4 P_1$, whereas the dashed pink line corresponds to $P = P_\infty = 0.77 P_1$. }
    \label{fig:G_colourmap_free_equator_dipole}
\end{figure}


\subsection{Non-fixed boundary condition at the equator}

In most previous studies that used classical methods to tackle this problem, it was a common practice to solve it in just one hemisphere and setting boundary conditions at the equator (Eq.~\eqref{eq:bc_equator}), where a kind of "jump" is expected to occur. This approach is reasonable, but it limits the range of magnetospheric solutions that can be obtained. The solutions have equatorial symmetry (in addition to axisymmetry) and end up having a specific configuration: a dipole magnetic field at the surface and a Y-point configuration where the equatorial and separatrix current sheets meet.

However, by utilizing PINNs and taking advantage of their local and flexible nature for imposing constraints and boundaries, we can create solutions with fewer (only physically relevant, rather than mathematical) requirements. In this section, we introduce solutions where boundary conditions are applied only at the surface and at infinity, without restricting the equator as part of the solution domain. 

At infinity, we simply demand that the solution approaches a split monopole configuration (last term in Eq. \eqref{eq:fb_parametrisation_free_equator}) with a specific value (denoted by $P=P_\infty$) at the equator:
$$
P(\mu, q=0) = P_\infty (1 - |\mu|)~.
$$
Since the equator is free from any boundary conditions, some other constraint must be imposed to distinguish between a possibly infinite class of different solutions. We decide to set the value of $P_c$ beforehand. This value separates the regions with and without electric currents. This approach is just as valid and perhaps more versatile than other constraints, like pinning down the position of the Y-point. As for $P_\infty$, we do not fix its value; instead we let the network figure it out.
The value of $P_\infty$ separates the regions with currents of different sign.

In Fig. \ref{fig:G_colourmap_free_equator_dipole} we present a typical solution for these boundary conditions.
This magnetospheric configuration bears resemblance to the one obtained in \citet{Contopoulos_2014}, where a substantial number of field lines that cross the light cylinder close inside the equatorial current sheet.
However, we have arrived to this result by following a different prescription. 
They enforced specific boundary conditions at the equator to ensure that the perpendicular component of the Lorentz force applied to the equatorial current sheet becomes zero. On the other hand, our approach involved imposing that the solution becomes a split monopole with a certain magnetic flux at infinity while leaving the equator unrestricted.

An interesting generalisation of this 
set of solutions involves applying non-dipolar surface boundary conditions. Instead of sticking to the basic dipole case, the surface magnetic field can be a combination of various magnetic multipoles \citep{Gralla_2016}. This option is not feasible in the classical solution because having contributions of even multipoles breaks the equatorial symmetry.
Nevertheless, our solver can handle this situation without any issues, because it is working throughout the entire domain.
As we move to significant distances from the star, we anticipate the multipole solution to gradually converge towards the classical dipole solution.

\begin{figure*}
    \begin{subfigure}[t]{\columnwidth}
        \centering
        \caption{}
        \includegraphics[width=\columnwidth]{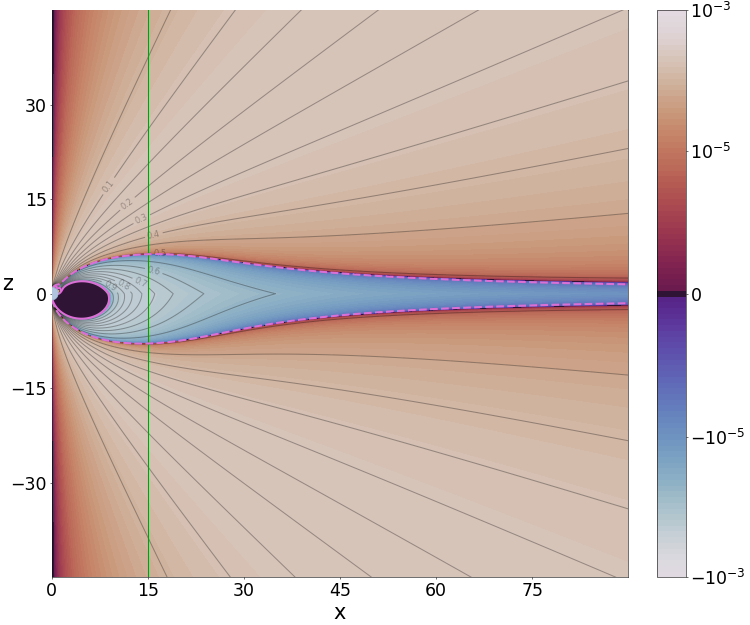}
    \end{subfigure}
    \begin{subfigure}[t]{\columnwidth}
        \centering
        \caption{}
        \includegraphics[width=\columnwidth]{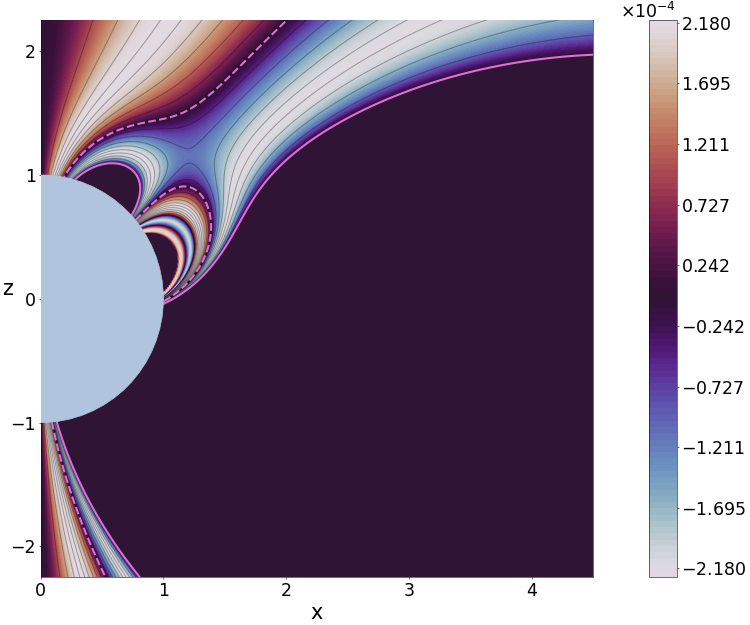}
    \end{subfigure}
    \caption{(\textbf{a}) Same as Fig. \ref{fig:G_colourmap_free_equator_dipole} but for a non-dipolar surface magnetic field. (\textbf{b}) Close up to highlight the multipolar content of the magnetic field close to the star. Notice that (\textbf{a}) is in logarithmic scale, while (\textbf{b}) is in linear scale.}
    \label{fig:G_colourmap_free_equator_multipole}
\end{figure*}

Fig. \ref{fig:G_colourmap_free_equator_multipole} shows an example of such a case. In particular, we consider a combination of a dipole, a quadrupole, and an octupole, with the corresponding coefficients in Eq.~\eqref{eq:fb_parametrisation_free_equator} being $b_1 = 1, b_2 = -2$ and $b_3 = 2$. At first glance, when observing the magnetosphere on a larger scale of tens of LC, it appears quite similar to the previous one. However, upon closer inspection of the zoomed-in right panel, the complex and intricate structure in the innermost region of the magnetosphere becomes evident. 

\begin{figure}
    \centering
    \includegraphics[width=\columnwidth]{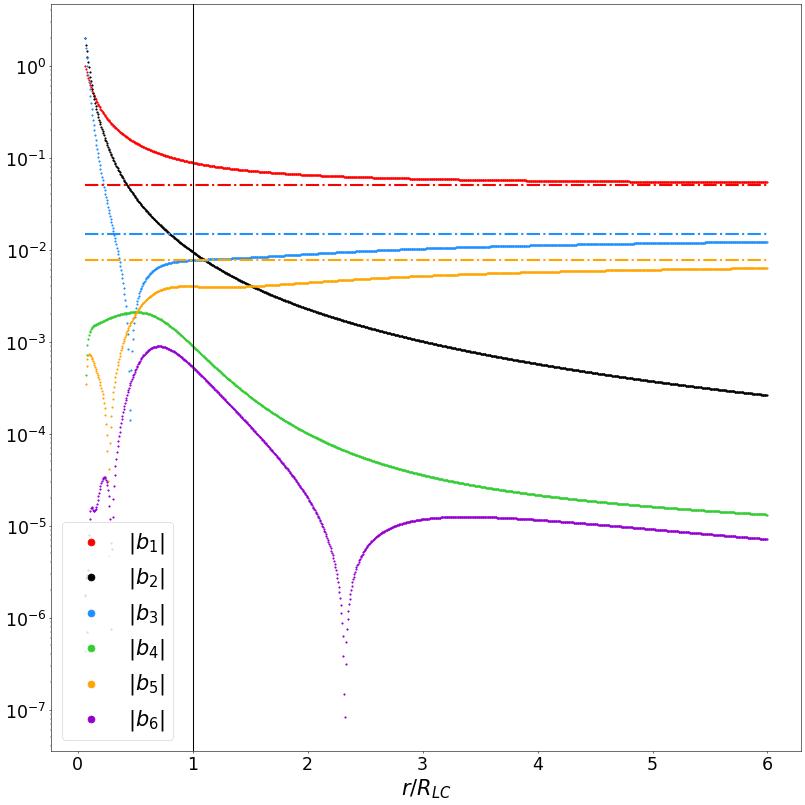}
    \caption{Coefficients $b_l$ of the $P$ polynomial expansion as a function of the radial distance $r$ for the non-dipolar case. Dashed dotted lines correspond to the even coefficients of the split monopole expansion, being zero the odd ones.}
    \label{fig:bl_non_dipole}
\end{figure}

To better understand the structure of this last model, in Fig. \ref{fig:bl_non_dipole}
we show the absolute value of the first six multipole weights
computed over spheres at different radii. We see how the even $l=2$ multipole quickly decreases with distance. Other even multipoles grow in the inner region, but they also become vanishingly small at a distance of a few LC. On the contrary, the odd multipoles approach the asymptotic values corresponding to the split monopole at radial infinity ($P \propto 1-|\mu|$), 
shown with dashed lines in the figure.

\section{Discussion and final remarks}

\begin{table}
    \begin{threeparttable}
        \caption{Model parameters for the different magnetospheric solutions}
        \label{tab:reference_solution_parameters}
        \begin{tabular}{lcccccccc}
            \hline
            Id. & $b_l$ & $\nicefrac{r_c}{\rlc}$  & $\delta P$ & $\nicefrac{P_c}{P_1}$ & $\nicefrac{P_\infty}{P_1}$ & $\Delta{\cal E}$ & $\dot{\cal E}$\\
            \hline
            C1 & 1,0,0 & 0.992 & 0.002 & 1.23 & - & 0.048 & 0.96  \\
            C2 & 1,0,0 & 0.7 & 0.002 & 1.71 & - & 0.089 & 1.91  \\
            C3 & 1,0,0 & 0.4 & 0.002 & 3.25 & - & 0.319 & 7.08 \\
            E1 & 1,0,0 & - & - & 1.40 & 0.77 & 0.024 & 0.39-0.73 \\
            E2 & 1,0,0 & - & - & 1.60 & 0.86 & 0.027 & 0.47-0.89 \\
            E3 & 1,0,0 & - & - & 1.80 & 0.95 & 0.031 & 0.57-1.10 \\ 
            E4 & 1,0,0 & - & - & 2.00 & 1.01  & 0.035 & 0.66-1.32  \\
            M1 & 1,-2,2 & - & - & 1.40 & 0.78 & 0.002\tnote{*} & 1.94-2.58 \\
            M2 & 1,-1,1 & - & - & 2.00 & 1.00 & 0.009\tnote{*} & 0.64-1.28 \\
            M3 & 1,-2,2 & - & - & 2.00 & 1.01 & 0.003\tnote{*} & 1.93-2.57\\
            M4 & 1,-3,3 & - & - & 2.00 & 1.01 & 0.002\tnote{*} & 1.94-2.58\\
            \hline
        \end{tabular}
        \begin{tablenotes}
           \item [*] for the cases with multipolar content, excess energy has been calculated w.r.t a non-rotating magnetic field with the same multipolar content ${\cal E}_{\text{mul}} = \frac{B_0^2 R^3}{2}\sum_{l}{b_l^2 \frac{l+1}{2l+1}}$.
        \end{tablenotes}
    \end{threeparttable}
\end{table}

Table \ref{tab:reference_solution_parameters} summarises the various models studied in this work. We use the letter $C$ to denote solutions acquired following the classical approach, the letter $E$ for solutions without an equatorial constraint and the letter $M$ for solutions with multipolar content. For the first three models, $P_c$ is determined by the PINN, while for the rest of them is fixed to a predetermined value. $P_\infty$, when present, is always determined by the PINN. 

The excess energy for all the models is of the order of a few percent. 
Models with no equatorial restriction tend to have lower energies, indicating that they are the preferred configuration in nature. With the exception of the model C3, the luminosity for all models does not vary from the classical dipole case by more than a factor of three.

\begin{figure}
    \centering
    \includegraphics[width=\columnwidth]{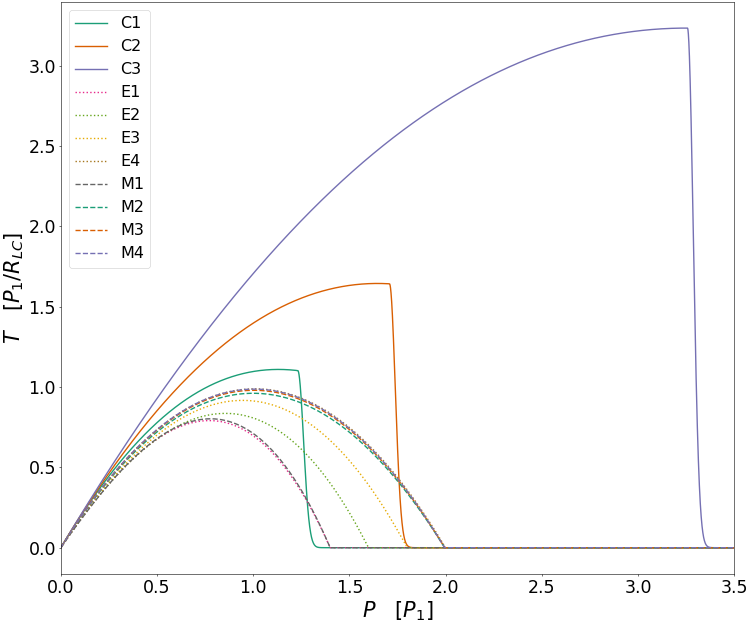}
    \caption{The functions $T (P)$ for the models of Table \ref{tab:reference_solution_parameters}.}
    \label{fig:T_of_P_comparison}
\end{figure}

In Fig. \ref{fig:T_of_P_comparison}, we present the function $T(P)$ for all the models of Table \ref{tab:reference_solution_parameters}.  
The curves corresponding to models with fixed boundary conditions at the equator exhibit a steep transition to zero, which is modelled by the current sheet (Eq.~\eqref{eq:current_sheet}) and begins at $P = P_c$.
In contrast, the models with no restrictions at the equator smoothly reach the value $P = P_c$
through a continuous transition from positive (orange area in Figs. \ref{fig:G_colourmap_free_equator_dipole}, \ref{fig:G_colourmap_free_equator_multipole}) to negative current (blue area in Figs. \ref{fig:G_colourmap_free_equator_dipole}, \ref{fig:G_colourmap_free_equator_multipole}).
Importantly, these models do not develop a current sheet to close the current circuit.

The area under each of the curves represents the luminosity (see Eq. \eqref{eq:luminocity}), which explains why the model with a lower $r_c$ exhibits a significantly higher energy loss rate, as it has more open lines carrying Poynting flux to infinity. Additionally, the presence or absence of equatorial restrictions leads to distinct behaviours in the models.
In the models without equatorial restrictions, not all the Poynting flux carried by lines crossing the light cylinder (LC) escapes to infinity. Instead, up to 50\% of the total pulsar spindown energy flux (the area between $P_\infty$ and $P_c$) remains confined in the equatorial current sheet. This trapped energy could potentially be dissipated in particle acceleration and high-energy electromagnetic radiation within a few times the light cylinder radius, as was first pointed out in \cite{Contopoulos_2014}. However, the exact fraction of power that can be locally dissipated and reabsorbed, as opposed to the fraction that is genuinely lost and contributes to the spin-down, remains unclear.
In the last column of Table \ref{tab:reference_solution_parameters} we have included the expected range of values for such models.

In this study, extending the results of our previous paper for slowly-rotating magnetar magnetospheres, our success lies in the proficient application of Physics-Informed Neural Networks (PINNs) to obtain numerical results for a diverse range of pulsar magnetospheric models, meticulously exploring various cases under different physical assumptions. Our extensive analysis encompasses the accurate reproduction of several distinct axisymmetric models found in the scientific literature. The decision to focus solely on axisymmetry was a deliberate one, aiming at affirming the remarkable ability of PINNs to effectively capture and characterize the intricate peculiarities exhibited by pulsar magnetospheres, including the features of the current sheet. Throughout our investigation, we purposefully accounted for a rich diversity of models, incorporating non-dipolar configurations. 

This work serves as a stepping stone towards the development of a robust and trustworthy general elliptic Partial Differential Equation solver, specifically tailored to address the challenging complexities of this and related astrophysical problems. Looking ahead, our research can be naturally extended to the three-dimensional magnetospheric case, a promising prospect that holds the potential for reaching a deeper understanding of the underlying physics governing pulsar magnetospheres. 

\section*{Acknowledgements}

We acknowledge the support through the grant PID2021-127495NB-I00 funded by MCIN/AEI/10.13039/501100011033 and by the European Union, the Astrophysics and High Energy Physics programme of the Generalitat Valenciana ASFAE/2022/026 funded by MCIN and the European Union NextGenerationEU (PRTR-C17.I1) and the Prometeo excellence programme grant CIPROM/2022/13. JFU is supported by the predoctoral fellowship UAFPU21-103 funded by the University of Alicante.

\section*{Data Availability}

All data produced in this work will be shared on reasonable request to the corresponding author.


\bibliographystyle{mnras}
\bibliography{Bibliography} 







\bsp	
\label{lastpage}
\end{document}